\documentclass[aps,amsmath,amssymb,prl,twocolumn,showpacs,superscriptaddress]{revtex4}
\usepackage{bm}
\usepackage{epsfig}
\usepackage[usenames]{color}
\usepackage{graphicx}
\usepackage{amsfonts}
\usepackage[hypertex]{hyperref}

\newcommand{\olcite}[1]{\cite{#1}} 

\renewcommand{\paragraph}[1]{\textit{#1.---} }

\renewcommand{\deg}{{^\circ}}

\begin{document}

\title{Nucleation of Spontaneous vortices in  trapped Fermi gases  undergoing a BCS-BEC crossover}

\author{A. Glatz}
\affiliation{Materials Science Division, Argonne National
Laboratory, 9700 S. Cass Av., Argonne, IL 60439, USA}

\author{H.~L.~L. Roberts}
\affiliation{James Franck Institute and Department of Physics,
University of Chicago, Chicago, IL 60637, USA}
\affiliation{Physics Department, University of California, Berkeley, CA 94720, USA}

\author{I.~S. Aranson}
\affiliation{Materials Science Division, Argonne National
Laboratory, 9700 S. Cass Av., Argonne, IL 60439, USA}

\author{K. Levin}
\affiliation{James Franck Institute and Department of Physics,
University of Chicago, Chicago, IL 60637, USA}

\date{\today }

\begin{abstract}
We study the spontaneous formation of vortices during the superfluid
condensation in a trapped fermionic
gas subjected to a rapid thermal quench via evaporative cooling.
Our work is based on the numerical solution  of
the time dependent crossover Ginzburg-Landau
equation coupled to the heat diffusion equation.
We quantify the evolution of condensate density and vortex length
as a function of a crossover phase parameter from BCS to BEC.
The more interesting phenomena occur somewhat nearer to the BEC
regime and should be experimentally observable;
during the propagation of the cold front,
the increase in condensate density
leads to the formation of supercurrents towards the center of the condensate
as well as
possible condensate volume oscillations.
\end{abstract}


\maketitle

\newpage

\paragraph{Introduction}
Atomic gases undergoing a superfluid condensation are very clean and simple
systems to address 
non-equilibrium quantum dynamics of superconductors.
As an example of important non-equilibrium behavior,
the present paper focuses on the way in which
a rapid quench from the normal phase leads to the formation of
a random tangle of vortices. In addition to cosmology \cite{Zurek}, 
the character and dynamics 
of these defects is of great
interest to a number of different physics subdisciplines ranging from atomic phy
sics,
\cite{weiler+n08,berloff+pra02} to condensed matter
\cite{bunkov,ruutu} and
high temperature superconductivity
\cite{Ong2,carmi,maniv}.
To this end, theoretical investigations of rapid quenches of atomic Bose gases
have been undertaken~\cite{weiler+n08} and appear
consistent with experiments. In this Rapid Communication
we make predictions for
future observations on 
{\it trapped Fermi} gases throughout the entire crossover from 
BCS to BEC, with special emphasis on the unitary regime.

These fermionic systems
are in many ways more suitable systems to study the dynamics of quantum gases.
The Pauli principle suppresses three body collisions making it possible to study
the strong interaction regime.  For the Bose counterparts, collisions are frequent
and lead to condensate decoherence.
A related advantage of the Fermi gases is the
opportunity to explore tunable
interaction strengths which, in turn, affects the dynamics. This tunability is
associated with the crossover between the
BCS (where the inter-fermionic attraction is weak)
and the Bose-Einstein condensed (BEC) limits (where the attraction is strong).
This crossover is entirely accessible
through application of a magnetic field, and the exploitation of so-called Feshbach resonances~\cite{Levinreview,Zwierlein}.
Our work is based on a numerical
simulation of  of
the time-dependent Ginzburg-Landau equation (TDGLE)~\cite{aranson+prl99,aranson+rmp02}
in the presence of thermal (white) noise $\chi$.
This TDGLE equation, which in many ways is one of the most fundamental
equations in condensed matter physics,
represents a differential equation characterizing the
dynamics and spatial dependence of the pairing gap parameter.  
Microscopically, one can demonstrate that
~\cite{Randeria,Maly,Levinreview}
the TDGLE coincides with the energy conserving
Gross-Pitaevski description in
the strong attraction or BEC limit, where the dissipation is minimal
Similarly in the weak attraction limit the TDGLE leads to
the well known diffusive dynamics of BCS theory. 

It should be stressed that TDGL theory refers to the gap or order
parameter, and not to the underlying fermions. In this way
the number of fermions, or the Fermi-Fermi interaction
parameters do not enter, as they have been effectively
integrated out. In contrast to
static GL theory, which is a consequence of Gorkov theory, there is no
fully rigorous derivation of this dynamics; alternative
dynamical schemes such as time dependent Bogoliubov deGennes
(TDBDG) theory \cite{BulgacScience}, which have similar antecedents
in Gorkov theory,
are also not rigorous.

By contrast to this TDBDG approach,
our studies are in the highly non-equilibrium regime.
We stress that there are essentially no alternative effective simulation techniques for addressing the condensate evolution, associated with
rapid quenches.
Monte Carlo and first principle quantum dynamics
are limited to small system sizes, short times and typically zero temperature; moreover, they deal with the
dynamics of single particles, not the collective dynamics of the order parameter (in the
strong coupling limit) which we consider here.

\paragraph{Theory}
We model the temperature equilibration in a physical way as a non-uniform process, associated with evaporative cooling.
The effects of the BCS-BEC crossover enter most prominently via the change in the complex
time relaxation rate
 of the TDGLE~\cite{Randeria,Maly,Levinreview}. 
As a function of the continuous crossover from BCS to BEC we see that the number and
lifetime of spontaneous vortices increases, so that the steady state condensate is
slower to form.
Among the most interesting observations in this paper pertains
to the near unitary regime (or mid-point between BCS and BEC) where the interactions are strongest and
the TDGLE- based approach is the most appropriate.

For notational purposes, it is convenient to introduce a parameter
$\theta=0\ldots\pi/2$ that {\it tunes} the equation from BCS ($\theta=0$) to BEC ($\theta=\pi/2$). Physically
 this parameter can be viewed as representing an applied magnetic field in conjunction with a Feshbach
 resonance. This tunes the strength of interaction between the fundamental fermionic particles (see e.g.~\cite{Randeria,Maly,Levinreview} for a rigorous derivation).

\begin{figure}[htb]
\includegraphics[width=.98\columnwidth]{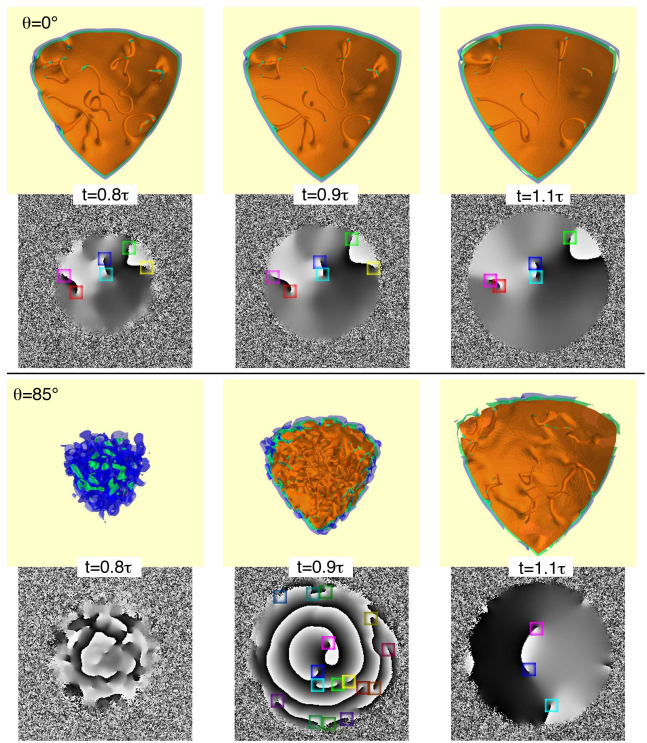}
\caption{(Color online) Isosurfaces and phase-cuts of the order parameter at different times. The upper row shows the 3D isosurfaces for the BCS ($\theta=0$) case at times $t=0.8\tau,0.9\tau,1.1\tau$, where $\tau$ is the  time needed to reach half of the condensate steady state volume [see Fig.~\ref{fig.tdep} (top)]. The  phase-cuts (phase $\varphi=\arg\psi$) are cuts through the center in the xy-plane - vortices are marked as the end-points of $2\pi$-phase jump lines (sharp black-white transition). The lower row shows the same pictures, but towards the BEC limit represented by  $\theta=85\deg$.}
\label{fig.orderparam}
\end{figure}

The TDGLE  is given by
\begin{equation}\label{eq.CGL}
e^{\imath \theta}\partial_t\psi=A[\mathbf{r},T(\mathbf{r},t)]\psi-\left|\psi\right|^2\psi+\frac{1}{2}\nabla^2\psi+\chi(\mathbf{r},t)
\end{equation}
where $\psi=\psi(\mathbf{r},t)$ is the order parameter and $\chi$ is uniformly distributed thermal noise with fluctuation temperature $T_{\chi}$.

The coefficient $A[\mathbf{r},T(\mathbf{r},t)]$ takes into account the evaporative cooling of the system and the trapping potential, i.e. $A[\mathbf{r},T(\mathbf{r},t)]=[1-T(r,t)]-U_0r^2$, where $U_0$ is the trapping potential which is determined by the trap size.
Eq.~(\ref{eq.CGL}) is written in dimensionless units, with length scale $\xi_0$, time scale $\tau_0$, and the order parameter scale $\psi_0$. Temperatures are measured in units of $T_c$. For (fermionic) Li$^6$ these values
are estimated as: $\xi_0=3.2\cdot 10^{-6}$m, $\tau_0=1.1\cdot 10^{-2}$s, and concentration $\psi^2_0=2.9\cdot
10^{16}$m$^{-3}$. The dimensionless fluctuation temperature for a condensate at $10$nK is in the interval $T_{\chi}\in [10^{-4};10^{-3}]$, depending on the condensed atoms.

The trap is assumed to be spherically-symmetric. Thus we calculate the temperature profile $T(r,t)$ due to evaporative cooling by solving the spherically-symmetric heat diffusion equation
\begin{equation}\label{eq.DT}
\partial_t T(r,t)=\mathcal{D}\left(\partial_r^2+\frac{2}{r}\partial_r\right)T(r,t)\,,
\end{equation}
with an initial uniform temperature $T(r<R,t=0)=T_0$ and final temperature $T(r\geq R,t)=T_f$ that is fixed at the boundary $r=R$. The  heat diffusion constant $\mathcal{D}$ is renormalized by $\xi_0^2/\tau_0$;  typical values for Li$^6$: $\mathcal{D}\simeq 3$.

All simulation runs start with random initial conditions in
the normal phase ($|\psi|^2\ll 1$) at temperatures $T_0$ larger than the critical temperature $T_c$. The evaporative cooling is
initiated at time $t=0$ on the surface of our spherical trap
with
radius $75\xi_0$. The cold front surface propagates
towards the center and quenches the atomic gas below the critical temperature.
The dimensionless diffusion constant was typically set to $\mathcal{D}=10$ and the fluctuation strength $T_{\chi}=10^{-4}$.
This mechanism of condensate nucleation is
to be contrasted with that described in Ref.~\olcite{weiler+n08} where the
cooling occurred uniformly in space, thus
corresponding to an infinite diffusion constant.
Another crucial difference with this previous work is the equilibration
mechanism: the time evolution of the condensate used in~\cite{weiler+n08}  was based on the
Gross-Pitaevskii model where the dissipation was introduced by truncation
of high-frequency modes.
By contrast, our TDGLE model includes a complex relaxation rate
which is determined by the crossover physics through
the phase factor $e^{\imath\theta}$.
Experiments on these cold gases \cite{Levinreview} cannot probe very deeply into the BCS or
BEC regimes, but are confined to the so-called ``unitary" midpoint.
Nevertheless, at unitarity, pairs are reasonably long lived~\cite{Maly} so that
we can associate $\theta \approx 70\deg ~-~ 85\deg$
with the physically accessible regime.

\paragraph{Numerical Results}
Our numerical calculations were done for  volumes discretized in up to $512^3$ grid points, averaged over up to $50$ different initial conditions and time evolved for $65$ different crossover phase values. The timestep in dimensionless units was chosen to be $0.1$ and the total simulation time up to $3000$ at $\theta$-values close to the BEC limit.
Our quasi-spectral split step method to solve the TDGLE which uses fast Fourier transforms, is much more stable than the traditional finite difference method.
As a result of employing modern graphics processing units, 
our systems are an order of magnitude larger than in
recent work on the bosonic BEC~\cite{weiler+n08}. This
allows us to simulate more realistic physical situations~\cite{CUDA}.

We begin with an illustration of the time evolved
condensation process for the BCS limit ($\theta =0$) and a near-BEC
situation ($\theta = 85\deg$).
Throughout this paper we avoid the strict fermion-based BEC limit ($\theta= \pi/2$)
since without dissipation
the condensate does not form and
nucleation of vortices is completely suppressed.
By introducing a complex relaxation rate in the TDGLE, we avoid
having to include
the interactions between condensed and
non-condensed pairs which do not enter as naturally, in the fermionic- BEC limit.
These were essential for addressing the bosonic counterpart experiments
\cite{weiler+n08,
bradley+pra08}.

Plotted in
Fig.~\ref{fig.orderparam} are the
isosurfaces for constant condensate magnitude. We also show
cross-sections in the $xy$-plane through the center of the system,
indicating the phase of the order parameter $\varphi=\arg\psi$. These latter
plots appear below the counterpart isosurface pictures and contain
information about spontaneous vortices.
The isosurfaces for the condensate magnitude are color-coded according to
fixed condensate density values $\rho_s\equiv |\psi|^2$, where we used $\rho_s=0.1,0.2,0.3$.
The three different panels from left to right correspond to three
different times close to the typical time $\tau$ at which the condensate forms, i.e. the time when the condensate volume reaches half of its saturation value.

One can see from
Fig.~\ref{fig.orderparam}, that most spontaneous vortices are connected to the surface of the condensate.
Their time evolution
is entirely accessible in our simulations~\cite{suppl} and one can follow their
decay  in the bulk and  at the condensate surface.
In the phase cross-section plots, these vortices are even more
visible appearing as topological defects (phase singularities).
Although vortices are most
plentiful at the surface of the condensate, we find their decay
occurs mostly in the bulk.

It is clear that in
the BCS limit the condensate forms quickly occupying the
entire trap volume in a relatively short time
period.
At much longer time scales (than shown here), the condensate
will appear uniform.
The near-BEC limit exhibits an interesting contrast:
Here the condensate expands slowly, and vortices are much
more plentiful, decaying even
more gradually.
An interesting feature is shown in the middle panel of the
last row of
Fig.~\ref{fig.orderparam}.
We observe in the course of condensate formation,
concentric rings in the phase cross-section interrupted by trapped topological defects.
This  corresponds to
the generation of supercurrents from the surface of the condensate towards its center.
The phenomenon of transient supercurrent $\mathbf{J}_s=\rho_s\nabla\varphi$ generation
can be understood as follows: In the BEC limit  the TDGLE can be written in the hydrodynamic form~\cite{landau_hf} with the condensate density $\rho_s$ satisfying the continuity equation $\partial\rho_s/\partial t=-\nabla \mathbf{J}_s$.
During the quench, $\rho_s$ changes from zero to its
maximal value  leading to the formation of supercurrents.

\begin{figure}[htb]
\includegraphics[width=\columnwidth]{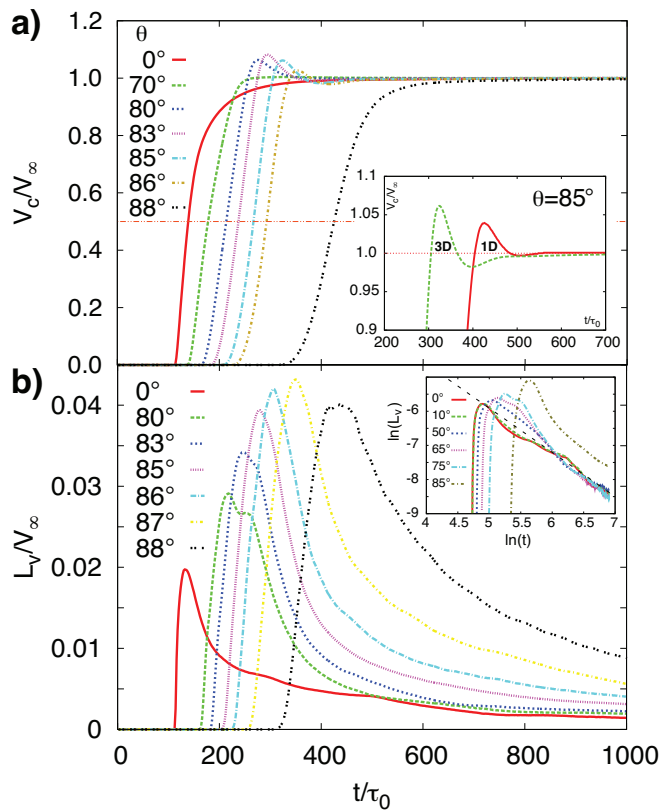}
\caption{(Color online) a) Condensate volume $V_c/V_{\infty}$ normalized to the asymptotic condensate volume $V_{\infty}$ with simulation time $t/\tau_0$  for different crossover phases $\theta$. The horizontal line at $V_c/V_{\infty}=1/2$ can be used to read the typical condensate formation time $\tau$. Inset:  the behavior of the condensate volume near the ``overshoot'' for $\theta=85\deg$ for the full simulation (labeled 3D) and the spherically-symmetric equation (labeled 1D).
b) Time evolution of the vortex length $L_v/V_{\infty}$ for different $\theta$. Inset:  a log-log representation illustrating a power-law decay for different $\theta$. The straight dashed line corresponds to a $1/t$ decay.
}
\label{fig.tdep}
\end{figure}

Figure~\ref{fig.tdep} presents a plot
of the time evolution of the condensate volume $V_c$, as well as
the vortex length $L_v$ for different values of the crossover phase $\theta$. Both quantities are normalized to the asymptotic final condensate volume.
Here the condensate volume is calculated from the number of grid points
with condensate density larger than $0.2$. Assuming a spherical shape of
the condensate (reflecting the trapping potential) we
establish the diameter of the condensate; we
then calculate the number of grid points with $\rho_s<0.2$ inside this sphere to estimate the vortex length.
In the BCS limit the dissipation is maximum which is responsible
for the rapid formation of the condensate. Increasing $\theta$
as one approaches the near-BEC gradually delays the formation
of the condensate leading to increase of  the maximum total vortex length.

The vortices show
a roughly $1/t$-decay of the total vortex length~\cite{berloff+pra02,Zurek}.
In Fig.~\ref{fig.tdep}b the time dependence of the total vortex
length is plotted
from the BCS to the near-BEC limit.
The inset shows some of these same curves on a log-log
scale and compares with a linear curve (dashed) representing a $1/t$ decay.
In two dimensions,  bulk vortices annihilate in pairs, the relaxation should be proportional
to the square of the vortex concentration, i.e.
the relaxation follows a power-law~\cite{vinen}.
Similarly, a $1/t$  is expected  for the decay of vortex lines  in 3D  for the BCS case
 and $1/t^{3/2} $ near the BEC limit~\cite{Kobayashi}.
The deviations from $1/t$ power law are likely caused by finite size effects.

\begin{figure}[htb]
\includegraphics[width=\columnwidth]{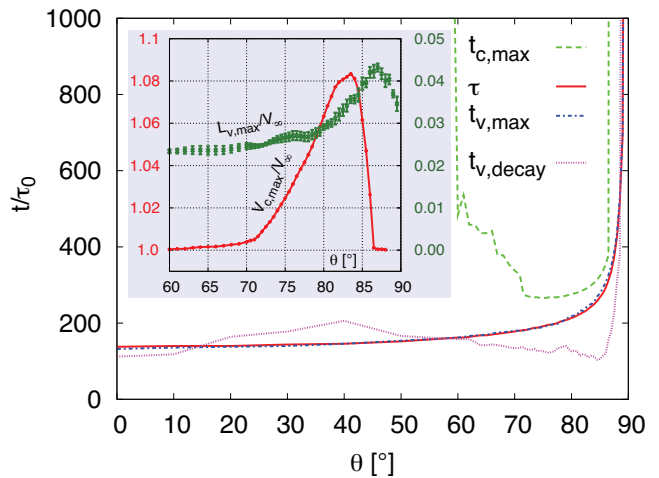}
\caption{(Color online)
Time scales extracted from the time evolution of the condensate volume and vortex
length [time in units of $\tau_0$]. $\tau$ is the time needed to reach half of the
 steady state condensate volume. $t_{c,\max}$ and $t_{v,\max}$ are the times when the
  condensate volume and vortex length has a local maximum respectively. $t_{v,\text{decay}}$ is the time over which the vortex length decays to $1/e$ of it maximum value.
Inset: Maximum of the volume of the condensate $V_{c,max}$ and  vortex length
 $L_{v,max}$ normalized to  $V_{\infty}$ versus
the crossover phase $\theta$. The data is thermally averaged over 10 to 50 realizations. }
\label{fig.overshot}
\end{figure}

The condensate volume
shows a well pronounced overshoot with subsequent
oscillations in time. This appears around unitarity, albeit closer to
the BEC limit. 
This overshoot is not generic, since it disappears in the 
BCS limit, presumably as a consequence of the large dissipation.
In
the near-BEC limit, where the dissipation is strongly suppressed, 
the condensate takes longer to form than the
time associated with a thermal quench.
We quantify this ``condensate breathing''  by studying
the height of this condensate volume overshoot in
the inset of Fig.~\ref{fig.overshot}. Plotted here is the maximum
of the volume of the condensate as a function of $\theta$. Also indicated is
the maximum vortex length
within the condensate.
It can be seen that
the maximum of the condensate overshoot occurs at $\theta=82.5\deg$.
The appearance of the overshoot is also characterized by the time scale $t_{\rm c,max}$, where the condensate reaches its maximum, which diverges below $\theta\sim 60\deg$ (since there is no overshoot)  and near the BEC limit (because the time of the occurrence of the overshoot   diverges for $\theta \to \pi/2$).
A minimum  of  $t_{\rm c,max}$ is observed at $75\deg$, i.e., the overshoot happens fastest.
Fig.~\ref{fig.overshot} indicates the typical time scale of the condensate formation $\tau$ and
the nearly identical time scale $t_{\rm v,max}$, which is the time at which the vortex length is maximum.
Both these times generally depend only weakly on $\theta$, except near $\theta \to \pi/2$.
Also the typical decay time of the vortices $t_{\rm v,decay}$ appears to be nearly independent of $\theta$ below $\sim 85\deg$ (see Fig.~\ref{fig.tdep}b).
Using the experimental time scales for Li$^6$, we can conclude from Figs.~\ref{fig.tdep} \&~\ref{fig.overshot},
that the condensate forms within a few seconds and vortices annihilate on the same time scale.

The overshoot and subsequent time oscillations of the condensate
may be related to excitation of sound waves produced
in a spherical trap by the quench.
To further clarify this phenomenon we solved the spherically-symmetric
TDGLE and obtained similar behavior
as shown in the inset to Fig.~\ref{fig.tdep}a; the main difference being a delay in the condensate nucleation~\cite{oneD}.
In the process we verified that the period of these oscillations is close to the time
needed for sound waves to traverse the condensate.
Indeed in our units the speed of sound
(which emerges from the BEC limit of the TDGLE)
is equal to unity in the long-wavelength
limit, while the oscillation period obtained from Fig.~\ref{fig.tdep}a is approximately $140$ time units. This
is close to the time
needed for the sound wave to cross our system with diameter $150$ length units. Using the  value for $\xi_0$ and $\tau_0$ for
Li$^6$, the speed of the second sound is on the order $10^{-4}$m/s, comparable to the mean velocity of the atoms at $10$nK.

There has been recent excitement over numerical approaches for addresses vortices
in Fermionic superfluids, particularly with the implementation \cite{BulgacScience} of a
time dependent Bogoliubov deGennes numerical scheme.
Unlike TDGLE, this approach
addresses the fermionic degrees of freedom in conjunction with
the
fermionic gap parameter. Generally it has been 
applied to $T=0$. Importantly, it does not incorporate the
fact that
the superconducting order parameter
need not be the same as fermionic excitation gap.
Related to this last observation is the fact that these schemes do not 
accomodate non-condensed bosons which are expected to enter
into the dynamics away from the strict BCS regime.
As a consequence, our understanding of vortices in fermionic superfluids
will require the exploration of a number of
different numerical approaches, including the TDGLE scheme we address here. 

We thank Kara Lamb, Matt Davis, Chih-Chun Chien and
Nate Gemelke for useful discussions. This work was supported by the
by the U.S. DOE, Office of Basic Energy Sciences, Division of Materials Science and Engineering, under Contract DEAC02-06CH11357,
and by NSF-MRSEC Grant No. 0820054 (KL).


\begin{thebibliography}{99}
\bibitem{oneD}
The spherically-symmetric TDGLE cannot capture the full dynamics and does not admit vortex solutions because of the reduced dimensionality.
Correspondingly the noise amplitude is reduced and becomes a function of the distance from the center of the condensate.
\end{thebibliography}
\end{document}